\documentclass{elsart1p} %% uses natbib so don't load cite or we get very mixed up
\usepackage{amsmath, amssymb, graphicx}

\begin{document}

\begin{frontmatter}
\date{}

\title{Implementing the LPM effect in a parton cascade model}
\author[duke]{C.E.Coleman-Smith}
\ead{cec24@phy.duke.edu}
\author[duke]{S.A.Bass}
\author[VECC]{D.K.Srivastava}

\address[duke]{Department of Physics, Duke University, Durham, NC 27701} 
\address[VECC]{Variable Energy Cyclotron Center, 1/AF Bidhan Nagar, Kolkata 700 064, India}
\begin{abstract}

Parton Cascade Models (PCM \cite{Geiger:1991nj, Bass:2002fh, Xu:2007aa, Molnar:2000jh}), which describe the full time-evolution of a system of quarks and gluons using pQCD interactions are ideally suited for the description of jet production, including the emission, evolution and energy-loss of the full parton shower in a hot and dense QCD medium. The Landau-Pomeranchuk-Migdal (LPM) effect \cite{LandauPom, PhysRev.103.1811}, the quantum interference of parton wave functions due to repeated scatterings against the background medium, is likely the dominant in-medium effect affecting jet suppression. We have implemented a probabilistic implementation of the LPM effect \cite{Zapp:2008af} within the PCM which can be validated against previously derived analytical calculations by Baier et al (BDMPS-Z) \cite{Baier:1996vi, Baier:1994bd, Baier:1996kr, zakharov-1, zakharov-2}. 

Presented at the 6th International Conference on Physics and Astrophysics of Quark Gluon Plasma (ICPAQGP 2010).

\end{abstract}


\end{frontmatter}

\section{Introduction}
Recent data from the Relativistic Heavy-Ion Collider (RHIC) at Brookhaven have provided strong evidence for the existence of a transient Quark-Gluon-Plasma (QGP) with the properties of a near ideal fluid and high opacity.
The discovery of the QGP has been anchored by the measurement of 
elliptic flow and the suppression of particles with high transverse momentum, jet quenching.

The suppression of high $p_T$ hadrons as measured at RHIC \cite{PhysRevLett.89.202301, PhysRevLett.88.022301} is a key signature of the formation of a QGP and of the modification of jets, collections of collimated high momentum particles, by this matter. The modification of hard probes provides an opportunity to learn about the physical properties of the QGP.

It has been shown that elastic energy loss is insufficient to entirely account for the fivefold suppression of high momentum partons as seen in the RHIC data \cite{PhysRevLett.88.022301}. Medium induced radiation must therefore play an important part in this process, furthermore at high probe energies quantum effects cannot be neglected \cite{Baier:2000mf, Majumder:2010qh}. The LPM effect is a quantum modification to the radiation of highly energetic particles passing through amorphous matter. At high energies forward scattering momentum transfers will be small and so the wavefunction of the particle will probe long distances in the material. Interactions with the medium can no longer be treated as single scatterings. Instead interactions with collections of scattering centers must be treated coherently.  This gives rise to interference effects which suppress the radiation spectrum.

In the non-Abelian LPM effect it is the rescattering of bremsstrahlung gluons with medium particles which leads to the dominant interference process. During emission collinear soft gluons are indistinguishable from their parent partons for some period, which can be characterized by a formation time $\tau_f = \omega / \mathbf{k}_{\perp}^2$. If this time becomes longer than the mean free path, i.e small emission angles and soft radiation, the gluon will interact coherently with additional scattering centers. As a result the radiation process becomes delocalized, with some characteristic number of scatterers being responsible for a single emission. This leads to a suppression of the radiation spectrum and a strong length dependence of the energy loss of the emitting parton \cite{Baier:1996vi, Baier:1996kr}. This effect is believed to provide the dominant contribution to the observed suppression of leading hadrons.

\section{The Parton Cascade}
We simulate the evolution of an infinite brick of partonic QGP matter with the VNI/BMS parton cascade model \cite{Geiger:1991nj, Bass:2002fh} a fully-relativistic Monte-Carlo transport code which numerically solves the Boltzmann equation,
\begin{equation}
\label{eqn-boltzmann}
p^{\mu} \frac{\partial}{\partial x^{\mu}} F_k(x, p) = \sum_{i}\mathcal{C}_i F_k(x,p).
\end{equation}
 The collision term $\mathcal{C}_i$ includes all possible $2 \to 2$ scatterings which may also produce final state radiation,
\begin{equation}
  \label{eqn-pcm-collision}
  \mathcal{C}_i F_k(x,\vec{p}) = \frac{(2 \pi)^4}{2 S_i} \int \prod_{j} d\Gamma_j | \mathcal{M}_i | ^2 \delta^4\left(P_{in} - P_{out}\right) D(F_k(x, \vec{p})),
\end{equation}
$d\Gamma_j$ is the Lorentz invariant phase space for the process $j$, $D$ is the collision flux factor and $S_i$ is a process dependent normalization factor.
A geometric interpretation of the total cross section for each process is used to determine which partons will interact, the partons are propagated along classical trajectories.

The PCM approach allows for the simulation of the evolution of a jet of fast partons and the reaction of the medium partons equally. The brick setup allows for a close examination of the energy loss processes free from the additional modifications due to motion through an expanding and cooling medium. This allows comparison with other theoretical models of jet energy loss \cite{vanLeeuwen:2010ti}. 

The medium is initialized with quarks and gluons sampled from thermal distributions typically at a temperature of 350MeV, the simulation volume has periodic boundary conditions to simulate an infinite brick of matter. A jet is created by injecting a hard parton with energy $\sim 100$GeV into the system, this leading parton is then tracked through the subsequent scatterings by tagging the hardest outgoing parton from each interaction. The strong coupling constant is fixed to $\alpha_s = 0.3$ for scatterings. The final-state radiation is produced in an angular ordered parton shower  whenever partons emerge from a collision with some time-like virtuality \cite{Webber:1986mc,Sjostrand:2006za}.

\section{Implementing the LPM effect}
Since the PCM is a Monte-Carlo process it is difficult to naturally include interference effects such as the LPM suppression. We implement the LPM effect using a local Monte-Carlo routine in the style of Zapp and Wiedemann \cite{Zapp:2008af}, which reproduces the leading BDMPS-Z \cite{Baier:1996vi, Baier:1994bd, Baier:1996kr, zakharov-1, zakharov-2} result for light-parton energy loss in a QGP medium $\Delta E \sim L^2$. This method is particularly appealing since it requires no artificial parametrization of the radiative process, it is a  purely probabilistic medium induced modification.

A shower of partons is produced by PYTHIA after an inelastic scattering. The hardest radiated gluon is selected to represent the shower as the probe and re-interact with the medium. This reflects the dominance of the gluon rescattering in the interference process. The formation times 
\begin{equation}
\label{eqn-formation-time}
\tau_f^{0} = \sum_{branchings}\frac{\omega}{\mathbf{k}_{\perp}^2},
\end{equation}
 for each branching during the parton shower leading up to the production of the probe gluon are summed. The probe parton is allowed to propagate through the medium and rescatter elastically during this time, the remainder of the partons from the radiation event propagate spatially but may not interact. Each time the probe gluon rescatters its formation time is recalculated as
 \begin{equation}
   \label{eqn-formation-time-recalc}
   \tau_f^{n} = \frac{\omega}{\left(\mathbf{k}_{\perp} + \sum_{i=1}^{n} \mathbf{q}_{\perp,i}\right)^2}. 
 \end{equation}
this simulates the emission of the shower from $n$ centers which transfer their momentum coherently. After this formation time expires the radiation is considered to have separated from the initiating particles and all partons may once again interact and radiate.

\section{Results}
\begin{figure}
  \begin{center}
  \includegraphics[width=0.9\textwidth]{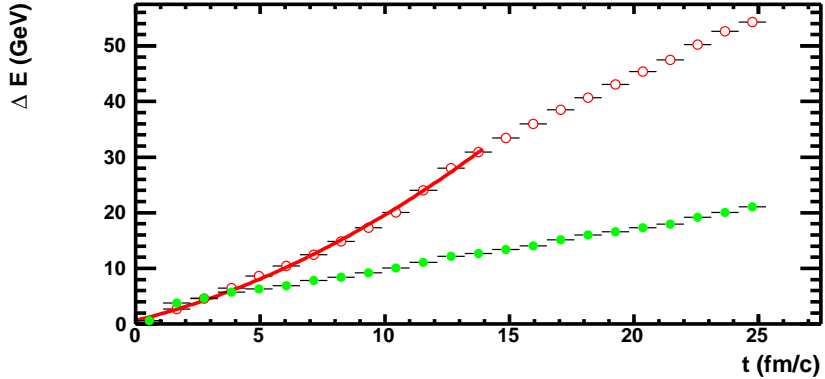}
  \caption{The total energy loss for a 100GeV light quark traversing a QGP medium at $T=0.35$GeV, with $\alpha_s = 0.3$. The open red circles show the energy loss from radiation with the LPM effect and elastic scattering, the closed green circles show the energy loss due to elastic scattering alone.} 
  \label{fig-main-result}
  \end{center}
\end{figure}

The total energy loss for a hard probe injected into the box is shown as a function of time in Fig: \ref{fig-main-result}. This energy loss is only incremented when the probe parton is outside of its formation time, when the parton is not in a coherent state with the emitted radiation.

The quadratic fit in Fig:\ref{fig-main-result} reproduces the leading order result from BDMPS \cite{Baier:1996kr}
\begin{equation}
  \label{eqn-bdmps-full}
  - \Delta E = \frac{\alpha_s C_A}{8} \frac{\mu^2}{\lambda_g} L^{2},
\end{equation}
for light partons the time elapsed and distance traveled through the box are interchangable and we use thermal values for the characteristic momentum transfer $\mu^2 = \frac{3}{2} g^2 T^2$ and the mean free path $\lambda_g^{-1} = 3TC_A \alpha_s \log{\frac{1}{\alpha_s}}.$ The elastic energy loss process has been found to agree well with theoretical predictions \cite{Shin:2010hu}.

We have successfully implemented a local Monte-Carlo version of the LPM effect and reproduced to leading order the canonical BDMPS result for the energy loss of the lead parton in a jet. The VNI/BMS parton cascade can now be used to examine how energy is carried away from the lead parton into the jet and the medium itself.

The authors would like to thank Berndt Mueller for many helpful discussions. We acknowledge support by DOE grants DE-FG02-05ER41367 and DE-SC0005396. This research was done using resources provided by the Open Science Grid, which is supported by the DOE and the NSF.

%% \bibliographystyle{elsarticle-num}
%% \bibliography{./lpm-proc.bib}{}

\begin{thebibliography}{10}
\expandafter\ifx\csname url\endcsname\relax
  \def\url#1{\texttt{#1}}\fi
\expandafter\ifx\csname urlprefix\endcsname\relax\def\urlprefix{URL }\fi
\expandafter\ifx\csname href\endcsname\relax
  \def\href#1#2{#2} \def\path#1{#1}\fi

\bibitem{Geiger:1991nj}
K.~Geiger, B.~Muller,  Nucl. Phys. B369 (1992) 600--654.

\bibitem{Bass:2002fh}
S.~A. Bass, B.~Muller, D.~K. Srivastava, Phys. Lett. B551 (2003) 277--283.

%\cite{Xu:2007aa}
\bibitem{Xu:2007aa}
  Z.~Xu and C.~Greiner,
  %``Transport rates and momentum isotropization of gluon matter in
  %ultrarelativistic heavy-ion collisions,''
  Phys.\ Rev.\  C {\bf 76}, 024911 (2007)
  %[arXiv:hep-ph/0703233].
  %%CITATION = PHRVA,C76,024911;%%

%\cite{Molnar:2000jh}
\bibitem{Molnar:2000jh}
  D.~Molnar and M.~Gyulassy,
  %``New solutions to covariant nonequilibrium dynamics,''
  Phys.\ Rev.\  C {\bf 62}, 054907 (2000)
  %[arXiv:nucl-th/0005051].
  %%CITATION = PHRVA,C62,054907;%%



\bibitem{LandauPom}
L.~D. Landau, I.~J. Pomeranchuk, Dolk. Akad. Nauk. SSSR 92~(92).

\bibitem{PhysRev.103.1811}
A.~B. Migdal, Phys. Rev. 103~(6) (1956) 1811--1820.

\bibitem{Zapp:2008af}
K.~Zapp, J.~Stachel, U.~A. Wiedemann, Phys. Rev. Lett. 103   (2009) 152302.


\bibitem{Baier:1996vi}
R.~Baier, Y.~L. Dokshitzer, A.~H. Mueller, S.~Peigne, D.~Schiff, Nucl. Phys. B478 (1996) 577--597.

\bibitem{Baier:1994bd}
R.~Baier, Y.~L. Dokshitzer, S.~Peigne, D.~Schiff, Phys. Lett. B345 (1995) 277--286.

\bibitem{Baier:1996kr}
R.~Baier, Y.~L. Dokshitzer, A.~H. Mueller, S.~Peigne, D.~Schiff, Nucl. Phys. B483 (1997) 291--320.

\bibitem{zakharov-1}
B.~Zakharov, JETP Lett. 63 (1996) 952--957.

\bibitem{zakharov-2}
B.~Zakharov, JETP Lett. 65 (1997) 615--620.

\bibitem{PhysRevLett.89.202301}
C.~Adler,  Phys. Rev. Lett. 89~(20) (2002) 202301.

\bibitem{PhysRevLett.88.022301}
K.~Adcox, Phys. Rev. Lett. 88~(2) (2001) 022301.

\bibitem{Baier:2000mf}
R.~Baier, D.~Schiff, B.~Zakharov, Ann.Rev.Nucl.Part.Sci. 50 (2000) 37--69.

\bibitem{Majumder:2010qh}
A.~Majumder, M.~van~Leeuwen, \href{http://arxiv.org/abs/hep-ph/1002.2206}
  {\path{arXiv:hep-ph/1002.2206}}.

\bibitem{Webber:1986mc}
B.~R. Webber,  Ann. Rev.  Nucl. Part. Sci. 36 (1986) 253--286.

\bibitem{Sjostrand:2006za}
T.~Sjostrand, S.~Mrenna, P.~Z. Skands, {PYTHIA 6.4 Physics and Manual}, JHEP
  0605 (2006) 026.

\bibitem{vanLeeuwen:2010ti}
M.~van Leeuwen, \href
  {http://arxiv.org/abs/1012.2261} {\path{arXiv:1012.2261}}.


\bibitem{Shin:2010hu}
G.~R. Shin, S.~A. Bass, B.~Muller, J. Phys. G37 (2010) 105112.




\end{thebibliography}

\end{document}